\documentstyle[aps,manuscript,epsfig,psfig]{revtex}
\newcommand{\be}{\begin{equation}}
\newcommand{\ee}{\end{equation}}
\def\beqa{\begin{eqnarray}}
\def\eeqa{\end{eqnarray}}

\def\lap{\lower.5ex\hbox{$\; \buildrel < \over \sim \;$}}
\def\gap{\lower.5ex\hbox{$\; \buildrel > \over \sim \;$}}

\tightenlines

\begin{document}
\preprint{
\vbox{   \hbox{CERN-TH/2001-140}
         \hbox{IFIC/01-29}
         \hbox{hep-ph/0105294}}}
\title{Status of the Gribov--Pontecorvo Solution to the Solar Neutrino Problem}
\author{V. Berezinsky$^{1}$, M.C. Gonzalez-Garcia$^{2,3}$ and 
C. Pe\~na-Garay $^{3}$}
\address{
$^{1}$ INFN, Laboratori Nazionali del Gran Sasso,
67010 Assergi (AQ) Italy\\
$^{2}$ Theory Division, CERN CH-1211, Geneva 23, Switzerland\\
$^{3}$Instituto de F\'{\i}sica Corpuscular,  
Universitat de  Val\`encia -- C.S.I.C\\
Edificio Institutos de Paterna, Apt 22085, 46071 Val\`encia, Spain}
\maketitle
\begin{abstract}
We discuss the status of the Gribov--Pontecorvo (GP) solution to the solar
neutrino problem. This solution naturally appears in bimaximal 
neutrino mixing and reduces the solar and atmospheric neutrino 
problems to vacuum oscillations of three active neutrinos.  
The GP solution predicts an energy-independent 
suppression of the solar neutrino flux. It is disfavoured by the rate of 
the Homestake detector, but its statistical significance greatly
improves, when the chlorine rate  and the boron neutrino flux are 
slightly rescaled, and when 
the Super-Kamiokande neutrino spectrum is included in
the analysis. Our results show that rescaling of the chlorine signal by only 
$10\%$ is sufficient for the GP solution to exist, if the
 boron--neutrino flux is
taken $10$ -- $20\%$ lower than the SSM prediction.
The regions allowed for the GP solution in the parameter space
are found and observational signatures of this solution are discussed. 
\end{abstract}
\section{Introduction}
Vacuum oscillations of maximally mixed $\nu_{\mu}$ and $\nu_{\tau}$ neutrinos 
with $(\Delta m^2)_{\rm atm} \sim 3\times 10^{-3}$~eV$^2$
are the favourite explanation of the atmospheric neutrino anomaly. A natural
generalization is bimaximal mixing \cite{BaPa}--\cite{Vis}
of three active neutrinos, when mixing is described  by the following 
matrix: 
\be
\left(
\begin{array}{c}
\nu_e \\ \nu_{\mu}\\ \nu_{\tau}
\end{array}
\right)
=
\left(
\begin{array}{ccc}
\frac{1}{\sqrt 2} & -\frac{1}{\sqrt 2} & 0\\
\frac{1}{2} & \frac{1}{2} & \frac{1}{\sqrt 2}\\
\frac{1}{2} & \frac{1}{2} & -\frac{1}{\sqrt 2}
\end{array}
\right)
\left(
\begin{array}{c}
\nu_1\\ \nu_2 \\ \nu_3
\end{array}
\right)\; .
\label{bimax}
\ee
In the case of the mixing matrix given by Eq.~(\ref{bimax}), the solar
neutrino oscillation is also maximal. To see this, it should be noted 
from Eq.~(\ref{bimax}) that  
\be
\frac{1}{\sqrt 2}(\nu_{\mu}-\nu_{\tau}) = \nu_3,
\ee 
while the other orthogonal combination of these states can be considered
as a new field
\be
\nu'=\frac{1}{\sqrt 2}( \nu_{\mu}+\nu_{\tau})\; .
\label{nu'}
\ee
Using the above equations, one obtains 
\be
\left(
\begin{array}{c}
\nu_e \\ \nu'\\ \nu_{\tau}
\end{array}
\right)
=
\left(
\begin{array}{ccc}
\frac{1}{\sqrt 2} & -\frac{1}{\sqrt 2} & 0\\
\frac{1}{\sqrt 2} & \frac{1}{\sqrt 2} &  0\\
\frac{1}{2} & \frac{1}{2} & \frac{1}{\sqrt 2}
\end{array}
\right)
\left(
\begin{array}{c}
\nu_1\\ \nu_2 \\ \nu_3
\end{array}
\right)\; .
\label{emax}
\ee
From Eq.~(\ref{emax}) it follows that $\nu_e$ and 
$\nu'=1/\sqrt{2}(\nu_{\mu}+\nu_{\tau})$ are a maximally mixed pair, 
and the flavour eigenstate $\nu_e$ is oscillating on the way from the Sun 
into $\nu'$, the coherent mixture of $\nu_{\mu}$ and $\nu_{\tau}$.

The above exercise is relevant to the Gribov--Pontecorvo (GP) \cite{GP} 
solution of the solar neutrino problem combined with atmospheric 
$\nu_{\mu}$--$\nu_{\tau}$ oscillations. Following \cite{GP}, the
definition of the GP solution can be given by two conditions: 
{\it (i)} smallness
of the oscillation length $l_{\nu}$ with respect to the mean distance
between the Sun and the Earth $\langle r\rangle= L$
\be
l_{\nu}= \frac{4\pi E}{\Delta m^2} \ll L=1.5\times 10^{13}~{\rm cm},
\label{oslength} 
\ee
and {\it (ii)} smallness of the matter corrections (MSW \cite{msw}) in 
the Sun and in the Earth \cite{earthmatter} 
(in Ref.\cite{GP} only vacuum oscillation is considered).

Indeed, in this case the averaged survival probability for $\nu_e$ is 
\be
\langle P_{ee}\rangle=1-\sin^2 2\theta\langle
\sin^2\frac{\Delta m^2 r}{4E}\rangle=
1-\frac{1}{2}\sin^2 2\theta \equiv P_{\rm GP} ,
\label{Pee}
\ee
and from comparison with experimental data, $P_{ee} \sim 0.5$, we come to
$\theta \sim \pi/4$, or to bimaximal mixing if 
$\nu_{\mu}\leftrightarrow \nu_{\tau}$ explains atmospheric neutrino
oscillations. 

Three remarks are immediately in order.

{\it(i)} There is no theoretical reason for bimaximal mixing to be 
exact, and
more generally one should consider {\it near-bimaximal mixing} 
\cite{BaPa,ourmaxmix}. 

{\it(ii)} The smallness of the matter correction effects, which we
included in the definition of the GP solution is actually not needed in the 
case of near-bimaximal mixing if only matter effects in the Sun are
included. For the exact maximal mixing,  
$\nu_e \to \nu'$ conversion in the Sun does not change the total
survival probability $\langle P_{ee}\rangle$ at the surface of the 
Earth \cite{BaGo,Giu}. 
In the arbitrary case where the mixing between $\nu_e$ and $\nu'$ 
is described by  the mixing angle $\theta$, and the MSW effect in the Sun 
converts $\nu_e$
into $\nu_S=\cos\theta_S \nu_1+\sin\theta_S\nu_2$ (subscript $S$ here
refers to the surface of the Sun), the total
survival probability $\langle P_{ee}\rangle =|\langle\nu_E|\nu_e\rangle|^2$ 
on the surface of the Earth can be readily calculated as 
\be
\langle P_{ee}\rangle=\cos^2\theta_S\cos^2\theta+\sin^2\theta_S\sin^2\theta
\label{Pe}
\ee
For exact bimaximal mixing $\cos^2\theta=\sin^2\theta=1/2$,  survival 
probability $\langle P_{ee}\rangle=1/2$ and thus it does not depend on 
 $\theta_S$, i.e. on how $\nu_e$ is 
converted in the Sun. For near-bimaximal mixing, 
Eq.~(\ref{Pe}) determines a narrow range of $\theta$ near $\pi/4$, 
where $\langle P_{ee}\rangle$ is practically energy independent,
i.e. it  does not
depend on $\theta_S$. The matter effect in the Earth, however, changes 
this conclusion as we shall see  in the next section.

\begin{figure}
\begin{center} 
\mbox{\epsfig{file=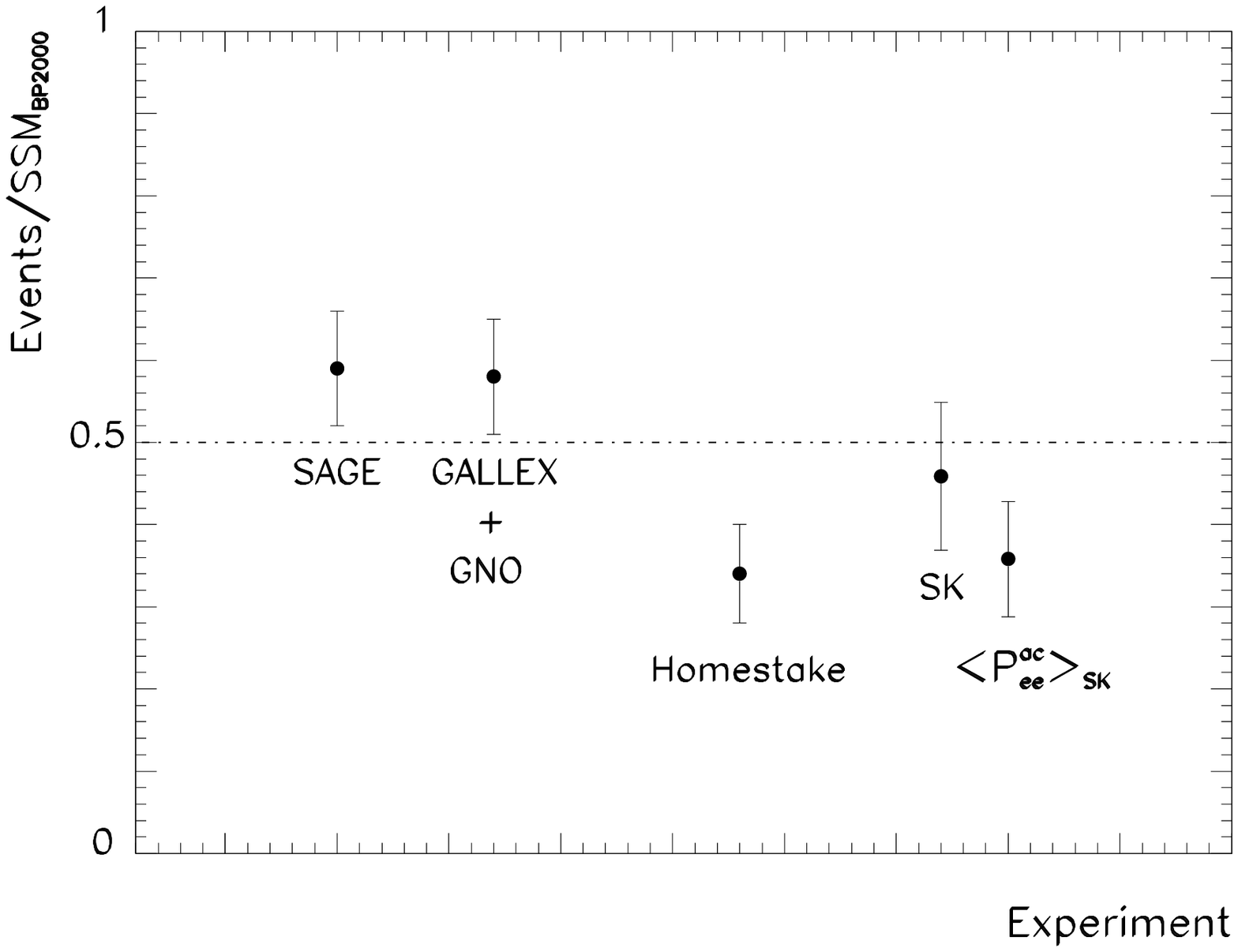,height=0.3\textheight}} 
\end{center} 
\caption{Ratios of observed rates to the BP00 prediction for 
the existing experiments. In the case of oscillations the ratios for 
all experiments, except Super-Kamiokande, are equal to the $\nu_e$ survival 
probability $P_{ee}$. For Super-Kamiokande the ratio SK gives $P_{ee}$ 
in the case of oscillation to sterile neutrinos, while for the
case of oscillation to active neutrino we plot 
the corresponding $P_{ee,ac}$.}
\label{fig_exp}
\end{figure}

{\it (iii)} The observational data (see Fig.~\ref{fig_exp}) do not support
$\langle P_{ee}\rangle$  being 
exactly energy-independent. While the recent Super-Kamiokande (SK) data 
agree well with $\langle P_{ee}\rangle$ being an energy-independent 
constant in the
energy interval 5 -- 14~MeV, the values $\langle P_{ee}\rangle$ 
from three different 
experiments, GALLEX--GNO/SAGE, Homestake and Kamiokande/SK,  
are not exactly the same.  

The aim of this paper is to discuss quantitatively the status of the GP
solution. An interesting region in the  parameter space is given by   
$\Delta m^2 \lesssim 10^{-3}~{\rm eV}^2$ and $\theta \sim \pi/4$, where 
bimaximal mixing is characterized by $(\Delta m^2)_{\rm atm}$ and 
$(\Delta m^2)_{\rm sol}$  not very different from each other, and where
the MSW effect is allowed as a small correction. 

Oscillations with energy-independent suppression were suggested
and studied in many works before \cite{AcLe,HaPe,Co,BaGo,Foot} and most
notably in the recent work of Ref.~\cite{choubey}. In all
these works the authors have realized that the observed rate in the
chlorine experiment (Homestake \cite{Hom}) contradicts
the energy-independent suppression, and it has to be taken larger than the
observed one for it to work. In this paper we argue that $10\%$ 
excess could be sufficient.
Note, that the solution with energy-independent
suppression is more general than the GP solution, because in the latter 
$l_{\nu}\ll L$ is assumed, while the energy-(quasi)independent solution
might appear in some other regions of the parameter space.

\section{Parameter space regions for the GP solution}

In this section we shall calculate the regions allowed for the GP solution
in the parameter space $\Delta m^2$, $\tan^2 \theta$. 
We first define the oscillation parameter space where the solar 
neutrino survival probability behaves effectively as the GP one.
In order to do so we impose the condition that for any of the
$i$ solar neutrino fluxes (integrated over the different production
point distributions) the survival probability in the relevant range 
of energies does not differ by 
more than 10\% (1\%) from $P_{\rm GP}$ given by  Eq.~(\ref{Pee}): 
 
\begin{equation}
\frac{|P_{ee}^i(E/\Delta m^2,\theta)-P_{\rm GP}(\theta)|}{P_{\rm GP}(\theta)} 
<0.1 \; (0.01) \;\;\;\;  \mbox{\rm for} \;\; E_{i,max}>E>E_{i,min},
\label{gpcond}
\end{equation}
where $E_{i,max}$ and $E_{i,min}$ determine the range of energies in which
the flux $i$ is detected in present experiments. For instance, for
$i=pp$, $E_{pp,min}=0.233$ MeV and $E_{pp,max}=0.42$ MeV. 
In the evaluation of the corresponding survival probabilities,
we have included the matter effects when propagating in the Sun and in the
Earth as well as the distance interference term:
\begin{equation} 
P_{ee}^i=P^S_{e,1}P^E_{1,e}+P^S_{e,2}P^E_{2,e}+
2\sqrt{P^S_{e,1}P^S_{e,2}P^E_{1,e}P^E_{2,e}}
\cos\frac{\Delta m^2_{21} (L)}{2E} \;, 
\label{P}
\end{equation}
where $P^S_{e,i}$ is the probability for the $\nu_e$ to exit the
Sun in the mass eigenstate $i$, while $P^E_{i,e}$ is the probability
for the mass eigenstate $i$ arriving at the Earth 
to reach the detector as a $\nu_e$. 
$L$ is the average distance between the Sun and the Earth.

\begin{figure}
\begin{center} 
\mbox{\epsfig{file=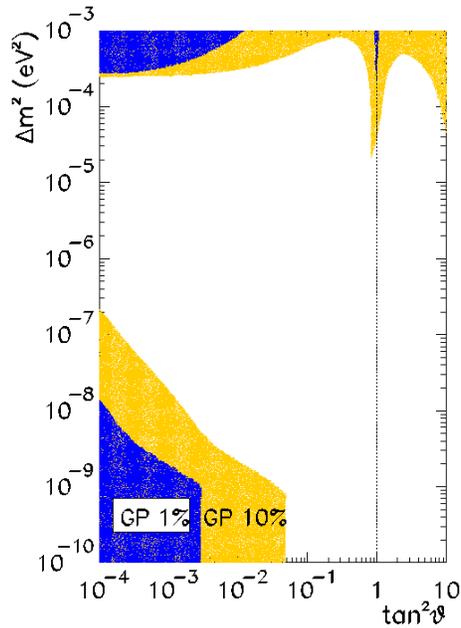,height=0.4\textheight}} 
\end{center} 
\caption{Regions in $\Delta m^2$, $\tan^2\theta$ parameter space, 
where the $\nu_e$ survival probability $P_{ee}(E)$ differs from the 
energy-independent GP survival probability $P_{\rm GP}$ by less than  10\% 
(lighter shadow) and less than 1\% (dark shadow). See the text for details. 
}
\label{regp}
\end{figure}

In Fig.~\ref{regp} we show the parameter space  $\Delta m^2$, 
$\tan^2\theta$ where the condition given by Eq.~(\ref{gpcond}) is verified
at 10\% (lighter shadow) and 1\% (darker shadow). 
The only interesting sector of the effective--GP region in this
parameter space is located at large $\Delta m^2$ around the maximal 
mixing line $\tan^2\theta=1$, where matter effects in the Sun are suppressed.  
This region is limited from above
by the CHOOZ reactor data \cite{chooz}. Only in this sector  
there is an overlap with the rate- and spectra-allowed regions (see below).

As was discussed above, for maximal mixing the matter effects in
the Sun do not alter the energy-independent survival probability 
$P_{ee}$ on the way from the production point inside the Sun to the
surface of the Earth. However Earth matter effects make $P_{ee}$ 
energy-dependent in the regions of maximal mixing at 
$10^{-5}\lesssim\Delta m^2\lesssim 10^{-8}$. In contrast with 
our calculations, 
the region $\Delta m^2\lesssim 10^{-7}$ is found in 
ref.\cite{choubey} as energy-independent one. We explain this
discrepancy by two effects: 
\begin{itemize}
\item For $10^{-8}\lesssim\Delta m^2\lesssim 10^{-7}$
Earth matter effects for $pp$-neutrinos result into an energy
dependence of the survival probability beyond 10\%. 
\item  At $\Delta m^2\lesssim 10^{-8}$
the $L$ dependent interference term in 
Eq.~(\ref{P}) gives strong energy dependence of the $^8$B flux. This
term was not included in the calculations of Ref.\cite{choubey}.  
\end{itemize}

\begin{figure}
\begin{center} 
\mbox{\epsfig{file=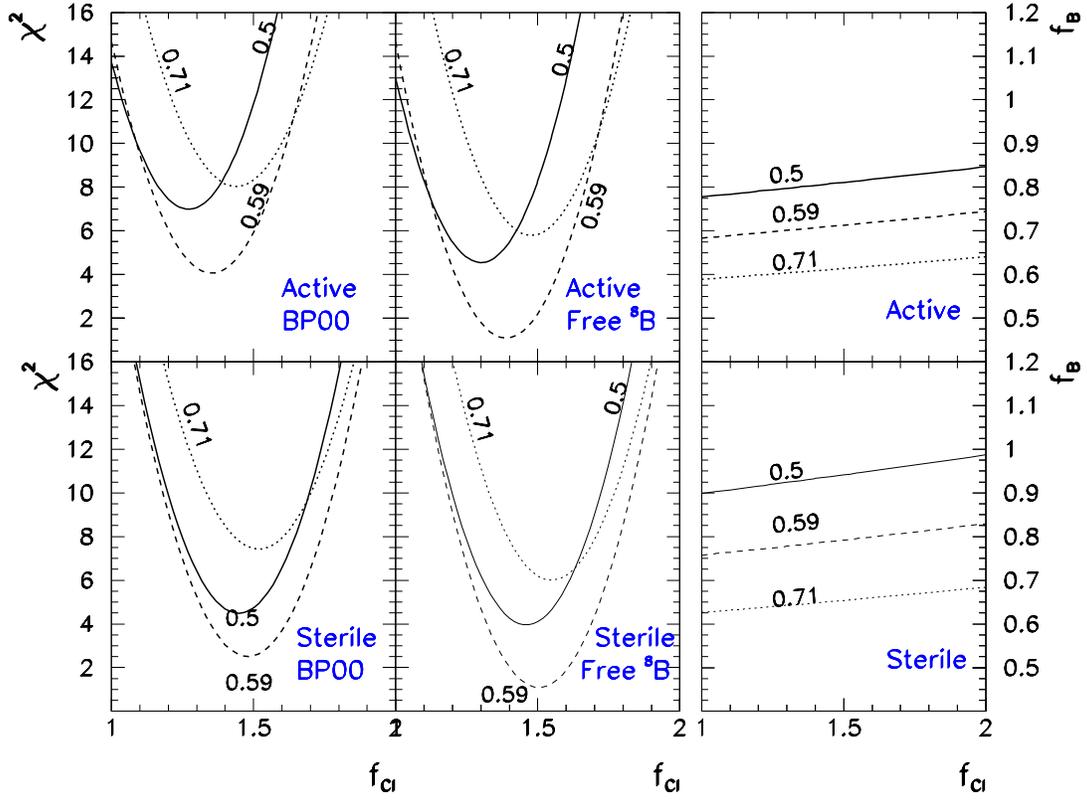,width=0.9\textwidth}} 
\end{center} 
\caption{ $\chi^2$ analysis of the three observed rates as a function of the 
$f_{\rm Cl}$ factor for different constant values of the survival
probability $P_{\rm GP}=0.5,0.59$ and $0.71$ (solid, dashed and dotted
lines respectively). The upper (lower) panels correspond to oscillations
into active (sterile) neutrinos. 
In the left panels we have used BP00 boron fluxes ($f_{\rm B}=1$), while in the
central panels $f_{\rm B}$ is left free to optimize $\chi^2$. The resulting 
$f_{\rm B}$ values which optimize $\chi^2$ for given $f_{\rm Cl}$ and
$P_{\rm GP}$ are plotted in the right panels.}
\label{chi2}
\end{figure}

As we mentioned above, the GP solution is incompatible with the central 
value of the rate measured by the Homestake detector $R_{\rm Cl}= 2.56$~SNU 
\cite{Hom}.
Following the prescription of many works, we shall use the rescaled rate 
$R_{\rm Cl}^{\rm res}=2.56f_{\rm Cl}$~SNU, assuming $f_{\rm Cl} \gtrsim 1$
to be a free parameter. In Fig.~\ref{chi2} we plot the $\chi^2$ function
from the analysis of the three observed rates as a function of the 
$f_{\rm Cl}$ factor for different constant values of the survival
probability $P_{\rm GP}=0.5,0.59$ and 0.71. 
The upper left panel corresponds to oscillations
into active neutrinos while the lower one into sterile neutrinos.
The differences between these two scenarios arise from the 
absence of NC contribution to the SK rate in case of oscillations 
into  sterile neutrinos.
From this figure we see that the best GP-like solution corresponds to
survival probability slightly larger than that for maximal--mixing case 
(close to 0.59) both for active and sterile neutrinos.  The quality
of these solutions are considerably improved when allowing a 30--50 \%
increase in  $f_{\rm Cl}$. This improvement is more significant
for the case of sterile neutrinos since the corresponding 
survival probability at SK
agrees better with the data from gallium detectors (see Fig.\ref{fig_exp}).

\begin{figure}
\begin{center} 
\mbox{\epsfig{file=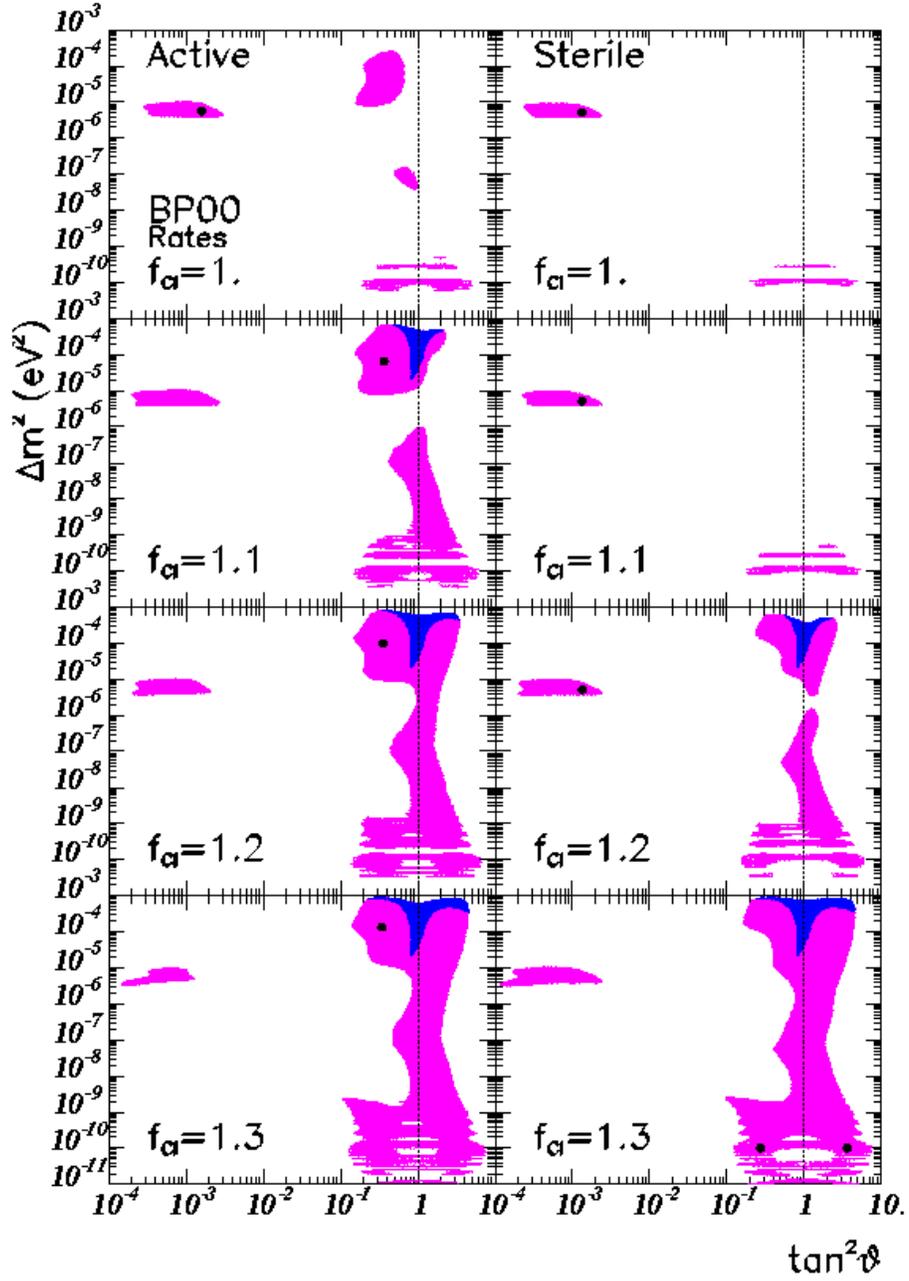,height=0.8\textheight}} 
\end{center} 
\caption{99\% CL regions allowed 
by the analysis of the experimental rates in GALLEX--GNO/SAGE, 
SK and Chlorine experiments, for
BP00 fluxes, for different 
values of $f_{\rm Cl}$ and for oscillations into active (left panels) and
sterile (right panels) neutrinos. 
The dots mark the position of the best-fit points  
in each panel. The effective GP solutions are marked as dark areas. }
\label{rates}
\end{figure}

This behaviour is also illustrated in 
Fig.~\ref{rates} 
where we show the $\Delta m^2$, $\tan^2\theta$ regions allowed 
by the statistical analysis of the rates of GALLEX--GNO/SAGE 
\cite{gallex,gno,sage}, SK \cite{sksol00} and
Chlorine\cite{Hom} experiments 
for different values of $f_{\rm Cl}$ in case of active and sterile 
neutrinos, and for Bahcall-Pinsonneault (BP00) \cite{BP00} 
fluxes.
The solutions, following the standard statistical analysis (for details see 
Ref.~\cite{oursolar}), are shown at 99\% CL. The 
effective GP solutions are marked as dark areas. Notice that
they appear at $f_{\rm Cl} \gtrsim 1.1$ (1.2) for active 
(sterile) oscillations  and that  all regions displayed 
 have a cut at $\Delta m^2\approx 8 \times 10^{-4}$ as a 
consequence of the CHOOZ \cite{chooz} bound.  

\begin{figure}
\begin{center} 
\mbox{\epsfig{file=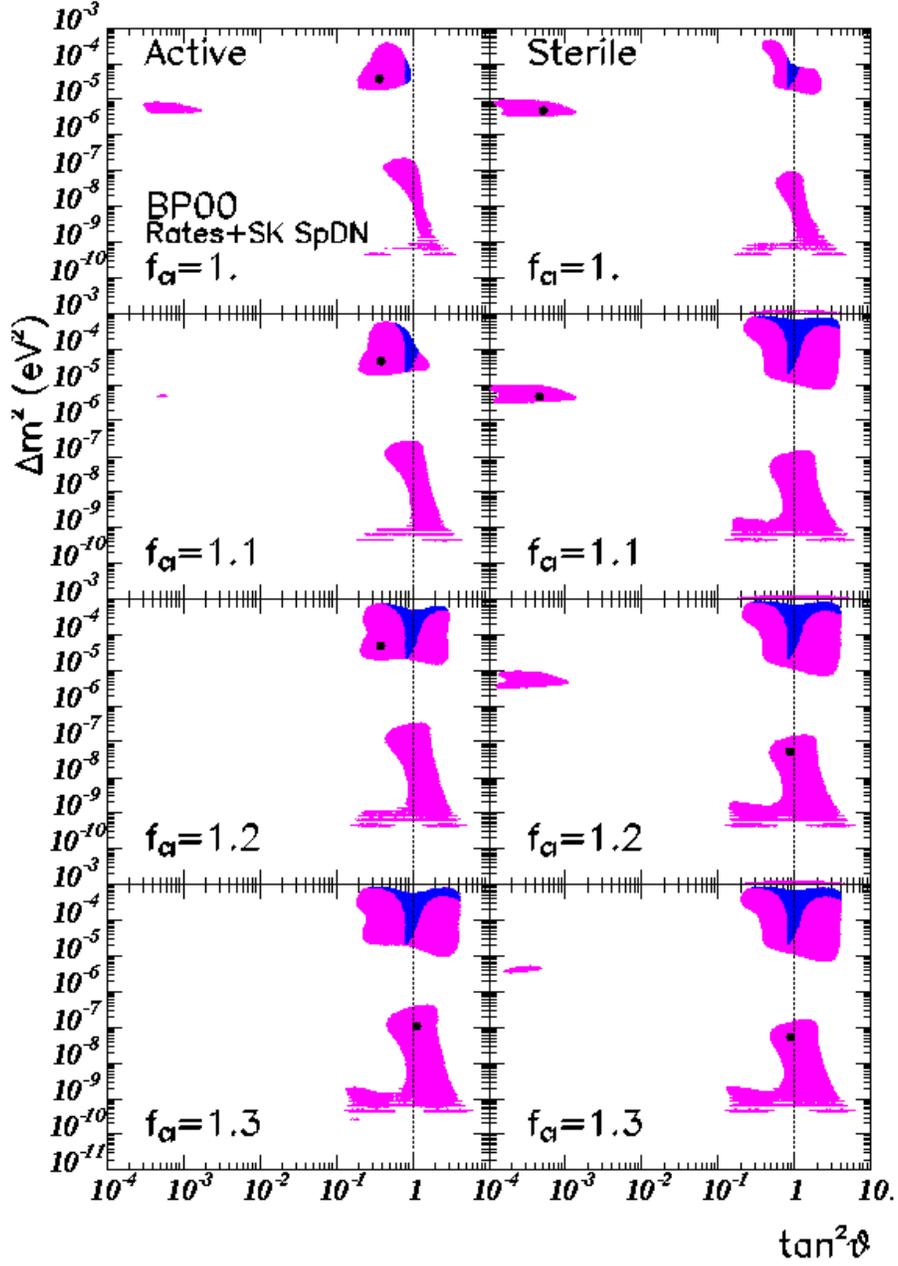,height=0.8\textheight}} 
\end{center} 
\caption{99\% CL regions allowed by the analysis of the three 
experimental rates and the SK day--night spectrum,  
for BP00 fluxes, for different 
values of $f_{\rm Cl}$ and for oscillations into 
active (left panels) 
and sterile (right panels) neutrinos. 
The dots mark the position of the best-fit points  
in each panel.
The effective GP solutions are marked as dark areas. }
\label{global}
\end{figure}

Inclusion of SK data on the energy spectra of boron neutrinos improves
the quality of the GP solution.
In Fig.~\ref{global} we display, for different values 
of $f_{\rm Cl}$,  
the regions allowed by the analysis of the rates and day--night spectrum of 
boron neutrinos measured by SK \cite{sksol00}.  Again the solutions
are shown at 99\% CL and the 
effective GP solutions are marked as dark areas. From these figures
one can see that the inclusion of the spectra data results in the appearance 
of allowed regions for the GP solutions at smaller values of $f_{\rm Cl}$. 
This is a natural result, because the rates 
of GALLEX--GNO/SAGE and Homestake can be also considered as information
about the solar neutrino spectrum, in its low energy part, however in 
contrast to the low energy part
of the spectrum, the GP solution describes well the spectrum observed in  
SK.

In Figs.~\ref{rates}--\ref{global} 
we have used the boron neutrino flux as 
calculated in the Standard Solar Model \cite{BP00}. This flux 
has a large theoretical uncertainty mostly due to uncertainties
in the pBe cross-section. In order to study the effect of a possible
deviation of the $^8$B flux from the SSM prediction \cite{BP00},
we shall introduce the rescaled boron neutrino flux defined as 
$\Phi_{\rm B}=f_{\rm B} \Phi_{\rm B}^{\rm SSM}$ with 
$\Phi_{\rm B}^{\rm SSM}$ given by \cite{BP00}.
In the central panels in Fig.~\ref{chi2}, we plot the $\chi^2$ function
from the analysis of the three observed rates as a function of  
$f_{\rm Cl}$ for different constant values of the survival
probability $P_{\rm GP}$ when the factor $f_{\rm B}$ is left free.
The upper central panel corresponds to oscillations
into active neutrinos, the lower one into sterile neutrinos.
In the right panels we show the corresponding values of  $f_{\rm B}$, 
which give the best agreement with the data for each value of
$f_{\rm Cl}$ and $P_{\rm GP}$. From this figure we find that 
although using a free $f_{\rm B}<1$  leads to a
further increase of the statistical significance  of the GP solution, it 
has a smaller
impact than the corresponding variation of $f_{\rm Cl}$. It, however,
allows for the presence of GP solutions with smaller $f_{\rm Cl}$. 
This is particularly the case for oscillations into active neutrinos.

\begin{figure}
\begin{center} 
\mbox{\epsfig{file=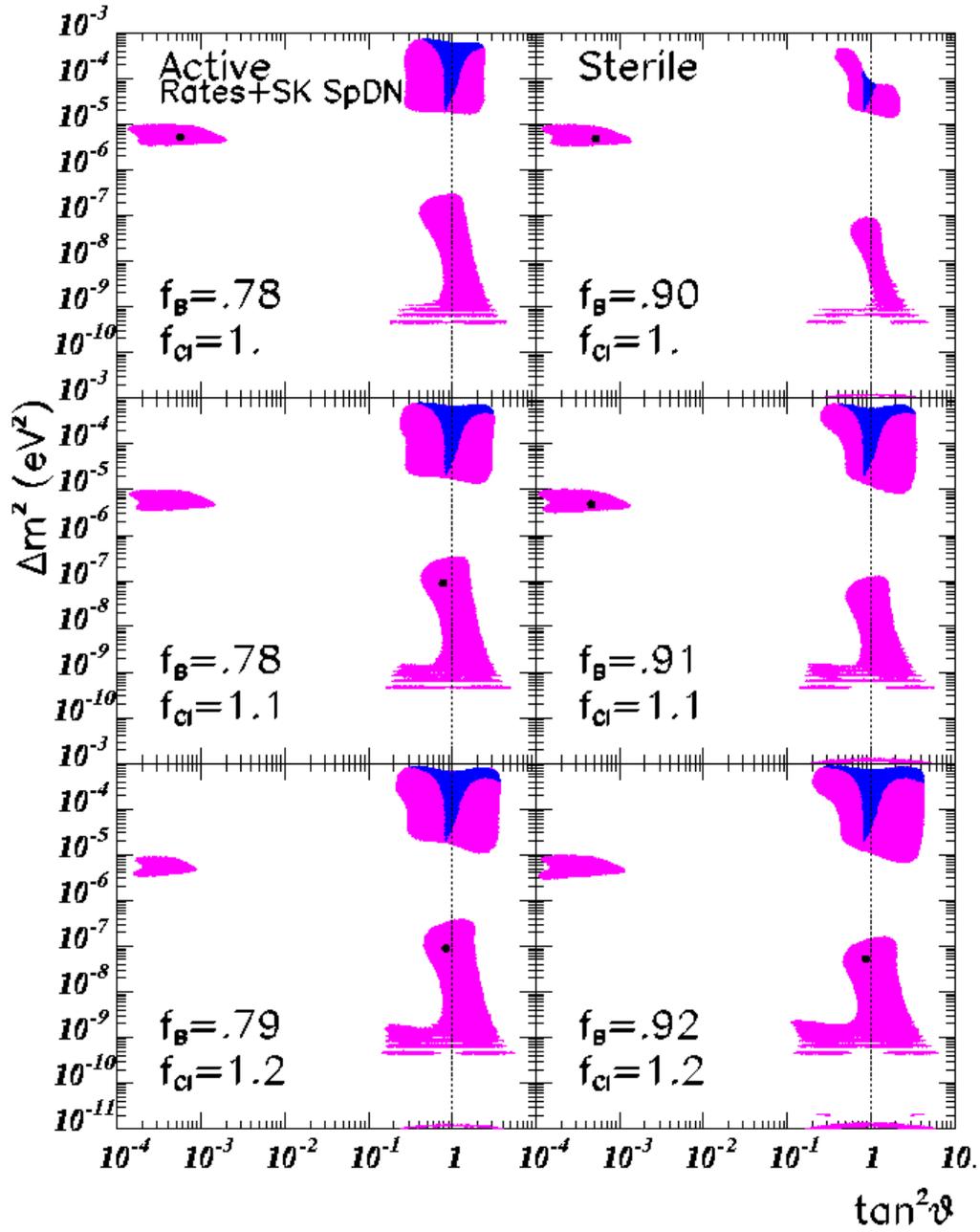,height=0.8\textheight}} 
\end{center} 
\caption{
99\% CL regions allowed 
by the analysis of the three experimental rates and the SK
day--night spectra for different 
values of $f_{\rm Cl}$  and $f_{\rm B}$, for oscillations into active
neutrinos (left panels) and sterile neutrinos (right panels). 
The effective  GP solutions are marked as dark areas. 
The dots mark the position of the best-fit points  
in each panel.}
\label{global_fb}
\end{figure}

In Fig.~\ref{global_fb} we plot the allowed
regions from the analysis of the rates and day--night spectrum of 
$^8$B  neutrinos measured by SK  for different values 
of $f_{\rm Cl}$ and $f_{\rm B}$. For the sake of concreteness we have 
chosen the $f_{\rm B}$ factor that gives a better fit to the three
rates for each value of $f_{\rm Cl}$ for maximal mixing. 
Namely, $f_{\rm B}$ is chosen as 
$f_{\rm B}\sim 0.8$ (0.9) for oscillations into active (sterile) neutrinos. 
In Fig.~\ref{global_fb} the
left panels correspond to oscillations into active neutrinos and
the right ones into sterile neutrinos. Comparing this figure with
the corresponding panels in Figs.~\ref{global} 
we see that lowering the $^8$B normalization leads to a larger overlap
between the allowed LMA region and the GP solution already for
$f_{\rm Cl}\leq 1.1$, in the case of  oscillations into active neutrinos. 

\section{Conclusions}
It could be that nature has chosen the most unsophisticated scheme 
of neutrino oscillations: three active neutrinos with 
(nearly) bimaximal mixing.  Mixing of $\nu_{\mu}$ and 
$\nu_{\tau}$ explains the atmospheric neutrino anomaly, and of $\nu_e$ and 
$\nu'=(1/\sqrt{2})(\nu_{\mu}+\nu_{\tau})$ the solar neutrino deficit.
In this case, the GP solution, provided by condition (\ref{oslength}), 
naturally  appears, and it is characterized by an energy-independent survival
probability $\langle P_{ee}\rangle= P_{\rm GP}$. 

The GP solution is disfavoured by the Homestake rate, but describes well 
the other rates as well as the energy spectrum observed in SK. 
The statistical significance of the GP solution strongly improves if one 
assumes rescaling of the  
chlorine rate by a factor $f_{\rm Cl}= 1.1$--$1.5$, while some 
further improvement arises if the $^8$B neutrino flux is also 
rescaled by a factor $f_{\rm B}=0.7$--$0.9$. 
In particular if the $^8$B flux happens
to be 10--20\% lower than the BP00-predicted central value, 
the GP solution for maximal mixing in active oscillations would be a\
llowed with  a chlorine rescaling factor $f_{\rm Cl}\lesssim 1.1$.

The GP solution will be directly searched for in the KamLand experiment 
\cite{Kam}. 
Detection of reactor neutrinos can result in the measurement 
of $\Delta m^2$ in the interval $10^{-3}$--$3\times 10^{-6}~{\rm eV}^2$
for large mixing angles. If $\Delta m^2$ is found outside the LMA MSW 
region or inside it at $\theta \approx \pi/4$, would mean the discovery of
the GP solution.

In low energy solar neutrino experiments the signatures of the GP solution
are the ordinary suppression of $^7$Be neutrinos 
given by a factor $P_{\rm GP}=\frac{1}{2}\sin^2 2\theta$
and the absence of anomalous
seasonal variations (beyond the geometrical ones). These features can be 
clearly seen in Borexino \cite{Bor} and KamLand \cite{Kam}
experiments. \\*[1mm]
\noindent{\bf Note}\\*[1mm]
This work was presented by M.C.Gonzalez-Garcia at the Gran Sasso Laboratory 
at the 5th Topical Workshop on 
``Solar Neutrinos: Where are the 
Oscillations?'' (March 2001). On March 29 the preprint 
by S.~Choubey, S.~Coswami, N.~Gupta and D.P.~Roy \cite{choubey}
appeared in the net. 
The basic assumptions they used, rescaling of the chlorine and boron 
fluxes, are the same as in our paper, but we are considering a GP
solution that is not, in principle, identical to the energy-independent 
solution, studied in the aforementioned paper.  The most noticeable 
difference in our results relates to the low $\Delta m^2$ solutions
found in Ref.\cite{choubey} and shown in their figures 1 -4. 
They are not present in our solutions partly due to interference term given
in our Eq.(\ref{P}) and disregarded in Ref.\cite{choubey}.

\section*{Acknowledgement}
We are grateful to Francesco Vissani for useful discussions. 
MCG-G is supported by the European Union Marie-Curie fellowship
HPMF-CT-2000-00516.
This work was also supported by the Spanish DGICYT under grants PB98-0693
and PB97-1261, by the Generalitat Valenciana under grant
GV99-3-1-01, and by the TMR network grant ERBFMRXCT960090 of the
European Union and ESF network 86.

\vskip 2cm
\vskip 2cm

\end{document}